\documentclass[11pt,twoside]{article}
\usepackage{asp2014}

\aspSuppressVolSlug
\resetcounters

\begin{document}

\title{iDaVIE-v: immersive Data Visualisation Interactive Explorer for volumetric rendering}

% full name: Lucia Marchetti
\author{Lucia~Marchetti$^{1,2,4}$, Thomas~H.~Jarrett$^{1,2}$, Angus~Comrie$^{1,2}$, Alexander~K.~Sivitilli$^{1,2}$, Fabio~Vitello$^4$, Ugo~Becciani$^5$ and A. R. Taylor$^{1,2,3}$}
\affil{$^1$Department of Astronomy, University of Cape Town, Private Bag X3, 7701 Rondebosch, Cape Town, South Africa; \email{lucia.marchetti@uct.ac.za}}
\affil{$^2$Inter-University Institute for Data Intensive Astronomy (IDIA), University of Cape Town, Cape Town, South Africa}
\affil{$^3$Department of Physics and Astronomy, University of the Western Cape, Private Bag X17, 7535 Bellville, Cape Town, South Africa}
\affil{$^4$INAF - Institute for Radio Astronomy, Via Gobetti 101, 40129, Bologna, Italy}
\affil{$^5$INAF - Catania Astrophysical Observatory, Via Santa Sofia 78, 95123, Catania, Italy}
% remove/add as you need

% remove/add authors as you need
\paperauthor{Lucia~Marchetti}{lucia.marchetti@uct.ac.za}{0000-0003-3948-7621}{University of Cape Town}{Astronomy Dept}{Cape Town}{Western Cape}{7701}{South Africa}
\paperauthor{Thomas~Jarrett}{tjarrett007@gmail.edu}{0000-0002-4939-734X}{University of Cape Town}{Astronomy Dept}{Rondebosch}{Western Cape}{7701}{South Africa}
\paperauthor{Angus~Comrie}{accomrie@gmail.com}{0000-0002-1790-0705}{University of Cape Town}{Astronomy Dept}{Cape Town}{Western Cape}{7701}{South Africa}
\paperauthor{Alexander~K.~Sivitilli}{alexandersivitilli@gmail.com}{}{University of Cape Town}{Astronomy Dept}{Cape Town}{Western Cape}{7701}{South Africa}
\paperauthor{Fabio~Vitello}{fabio.vitello@inaf.it}{0000-0003-2203-3797}{INAF}{Institute of Radioastronomy}{}{Bologna}{40129}{Italy}
\paperauthor{Ugo~Becciani}{ugo.beccani@inaf.it}{0000-0002-4389-8688}{INAF}{Catania Astrophysical Observatory}{}{Catania}{95123}{Italy}
\paperauthor{A. R.~Taylor}{russ@idia.ac.za}{0000-0001-9885-0676}{University of Cape Town}{Astronomy Dept}{Cape Town}{Western Cape}{7701}{South Africa}
% remove/add as you need

% leave these next few aindex lines commented for the editors to enable them. Use Aindex.py to generate them for yourself.
% first presenting author should be the first entry for bold-facing the author index page-reference
%\aindex{Marchetti,~L.}
%\aindex{Author2,~S.}
% remove/add as you need

% leave the ssindex lines commented for the editors to enable them, use Index.py to suggest yours
%\ssindex{FOOBAR!conference!ADASS 2020}
%\ssindex{FOOBAR!organisations!ASP}

% leave the ooindex lines commented for the editors to enable them, use ascl.py to suggest yours
%\ooindex{FOOBAR, ascl:1101.010}
  
\begin{abstract}
We present the beta release of \texttt{iDaVIE-v}, a new Virtual Reality software for data cube exploration. The beta release of \texttt{iDaVIE-v} (immersive Data Visualisation Interactive Explorer for volumetric rendering) is planned for release in early 2021. \texttt{iDaVIE-v} has been developed through the Unity game engine using the SteamVR plugin and is compatible with all commercial headsets. It allows the visualization, exploration and interaction of data for scientific analysis. Originally developed to serve the H{\textsc i} Radio Astronomy community for H{\textsc i} source identification, the software has now completed the alpha testing phase and is already showing capabilities that will serve the broader astronomical community and more. \texttt{iDaVIE-v} has been developed at the \textit{IDIA Visualisation Lab} (IVL) based at the University of Cape Town in collaboration with the \textit{Italian National Institute for Astrophysics} (INAF) in Catania.
\end{abstract}

\section{Introduction}
Virtual reality (VR) tools are the best suited for 3D (or multi-dimensional) data exploration. They enable a unique and immersive perspective on the data and allow intuitive interactions with the data. They thus speed up both the data interrogation process and the scientific discoveries that arose from it. Nevertheless, even though VR is widely developed and exploited by the gaming industry, it is only in its early days for scientific exploitation and is mostly used for science education or communication. 

The \texttt{iDaVIE-v} tool is part of the IVL\footnote{https://vislab.idia.ac.za/} \texttt{iDaVIE} software suite and has been developed in collaboration with INAF-Catania. It is the first VR tool developed by a team of developers and professional astronomers for scientific data analysis. \texttt{iDaVIE-v}'s primary scientific driver is the interrogation of Neutral Hydrogen (H{\textsc i}) Radio Data Cubes, where it aims to enable a set of critical operations on the data that are best carried out in an immersive environment. These operations are, for example, H{\textsc i} source detection, source identification, source characterisation and fast validation of (semi)automatic source extraction algorithms such as \texttt{SoFIA}\footnote{https://github.com/SoFiA-Admin/SoFiA}.
Even though it has been developed with these clear goals, its usage can be broader. The tool is flexible and can thus serve any science case that shares similar aims and uses similar data formats, such as the interrogation of medical and biological 3D datasets. 

In this paper we illustrate the main capabilities of \texttt{iDaVIE-v} in the astronomical context. For more details on the technical software development and broader reach of \texttt{iDaVIE-v} and of the complete software suite \texttt{iDaVIE} we refer the reader to \citet{P4-6adassxxviii}, \citet{O5-5adassxxviii} and \citet{Jarrett2020}.%Jarrett et al. 2021 (). 

\section{\texttt{iDaVIE-v} hardware requirements}
\texttt{iDaVIE-v} has been developed and tested to work with any commercial VR headsets currently available (e.g. Oculus Rift and Rift S, HTC Vive and Vive Pro or the Samsung Odyssey) and to (eventually) run on different operating systems (e.g. Microsoft Windows, Linux and MacOS). Nevertheless, the beta version presented here only runs on Microsoft Windows (Windows 10, version 1903 or newer). The minimum and recommended hardware requirements to run \texttt{iDaVIE-v} and avoid any efficiency issues are as follows: 
\begin{itemize}
    \item \textbf{Minimum requirements.} CPU: Quad core AMD Ryzen or Intel i5; Memory: 16 GB; Disk: SSD highly recommended; GPU: NVIDIA 1060 / NVIDIA 1650 Super / AMD Radeon RX 5500 XT or higher.
    \item \textbf{Recommended requirements.} CPU: AMD Ryzen R7 or Intel i9; Memory: 32 GB; Disk: NVMe SSD; GPU: NVIDIA 2070 / AMD Radeon RX 5700 XT or higher.
\end{itemize}

\section{\texttt{iDaVIE-v} capabilities for H{\textsc i} data cube studies}
To ease the understanding of the tool, in this section we briefly summarise the operations and actions that are currently enabled in the \texttt{iDaVIE-v} beta release and that are thought to serve the Radio H{\textsc i} astronomical community and its science needs. Any example reported hereafter will thus refer to this specific science case; for a more extensive description on the various scientific applications of the tool we refer the reader to \citet{Jarrett2020}%Jarrett et al. (in prep.).
\begin{itemize}
\item \textbf{Data import: the desktop GUI.} The user can load the data in \texttt{iDaVie-v} using an \textit{ad-hoc} developed desktop GUI. This is because the most common desktop operations such as the search of files, scrolling of lists, click buttons and write text are best performed using a keyboard outside the VR environment. The typical set of data that needs to be imported for astronomical H{\textsc i} science are the H{\textsc i} data cube of interest (in \texttt{fits} format) and, if available (optional), a mask (aka a cube with the same size and format of the data cube, but that contains only the H{\textsc i} source placeholders identified by a source finding algorithm like e.g. \texttt{SoFIA}) and a source catalogue (of any kind). The GUI (Fig. \ref{fig:plotGUI}) allows to browse the local files, to load the mentioned data and to inspect them.
\articlefigure[width=0.85\textwidth]{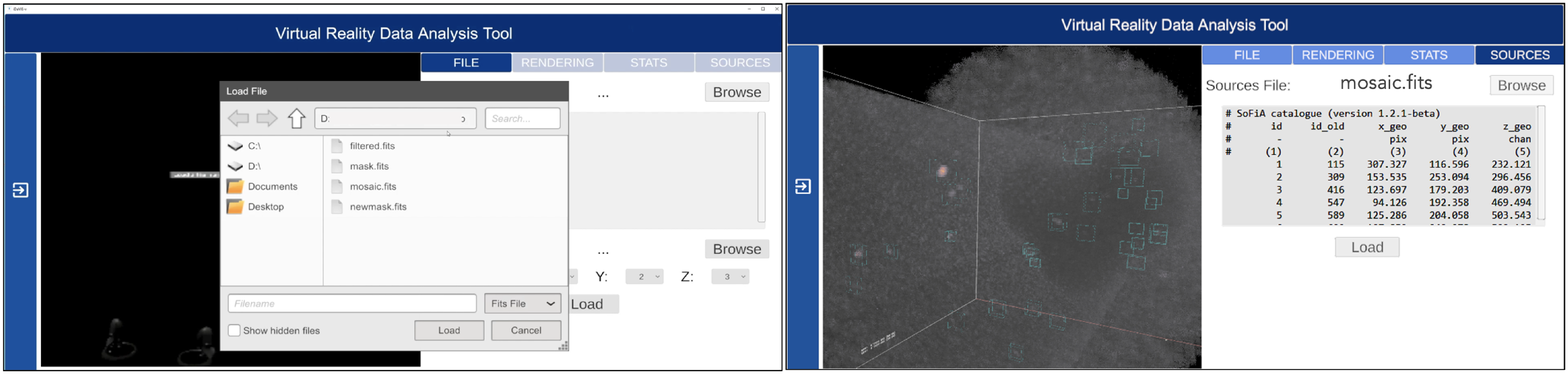}{fig:plotGUI}{\texttt{iDaVIE-v} Desktop GUI. \textbf{Left:} the interface to load the data. \textbf{Right:} feedback after ingestion of all input files (cube, mask and data catalogue) is completed. In both images the "black" square on the left shows the VR view (aka what the user sees when wearing the VR headset), while the space on the right shows the files headers and allows some operations such as select the colour transform, change the data visualisations thresholds, visualise some basics statistics of the data. Most of these operations can also be performed in the VR environment.}

\item \textbf{Data interaction in the VR space.} Once the data are loaded the user puts on the headset and hand controllers for immersive scientific analysis.
%can immerse in them and start the scientific analysis by wearing the VR headset and using the hand controllers.
\texttt{iDaVie-v} is designed and developed with the user interactions foremost in mind. By design the hand controllers of most commercial VR systems have similar limitations when it comes to menu based operations, for this reason, in \texttt{iDaVie-v} most of the data interaction operations can be performed both through menus, and/or through voice-activated commands coupled with 
a series of simple gestures (such as moving the controllers up and down or left and right to e.g. change thresholds). 
%and voice activated commands. 
In Fig. \ref{fig:plotVR} we show some of \texttt{iDaVie-v} capabilities and visualisation modes. 
The most desired functionalities by the H{\textsc i} astronomical community have been developed and implemented in the beta release. In a nuthshell, the user can: \textbf{a)} visualise the entire data cube or select and visualise only a portion of it for better analysis; \textbf{b)} move/rotate the data in any direction (or simply walk through the data)  and zoom in/zoom out on the data; \textbf{c)} change the colour transform to better highlight particular features in the data; \textbf{d)} overlay catalogues on the data cube; \textbf{e)} overlay a mask on the data cube; \textbf{f)} subtract the signal of the masked sources from the cube and explore the residuals; \textbf{g)} derive  in  real  time,  sky  and  source  statistics,  moment maps and other analytics that may be derived from the data and masks; \textbf{h)} edit the mask in real time by simply adding or subtracting voxels\footnote{3D volumetric pixels} to it; \textbf{i)} take VR-view snapshots that saves jpeg files.
\articlefigure[width=0.85\textwidth]{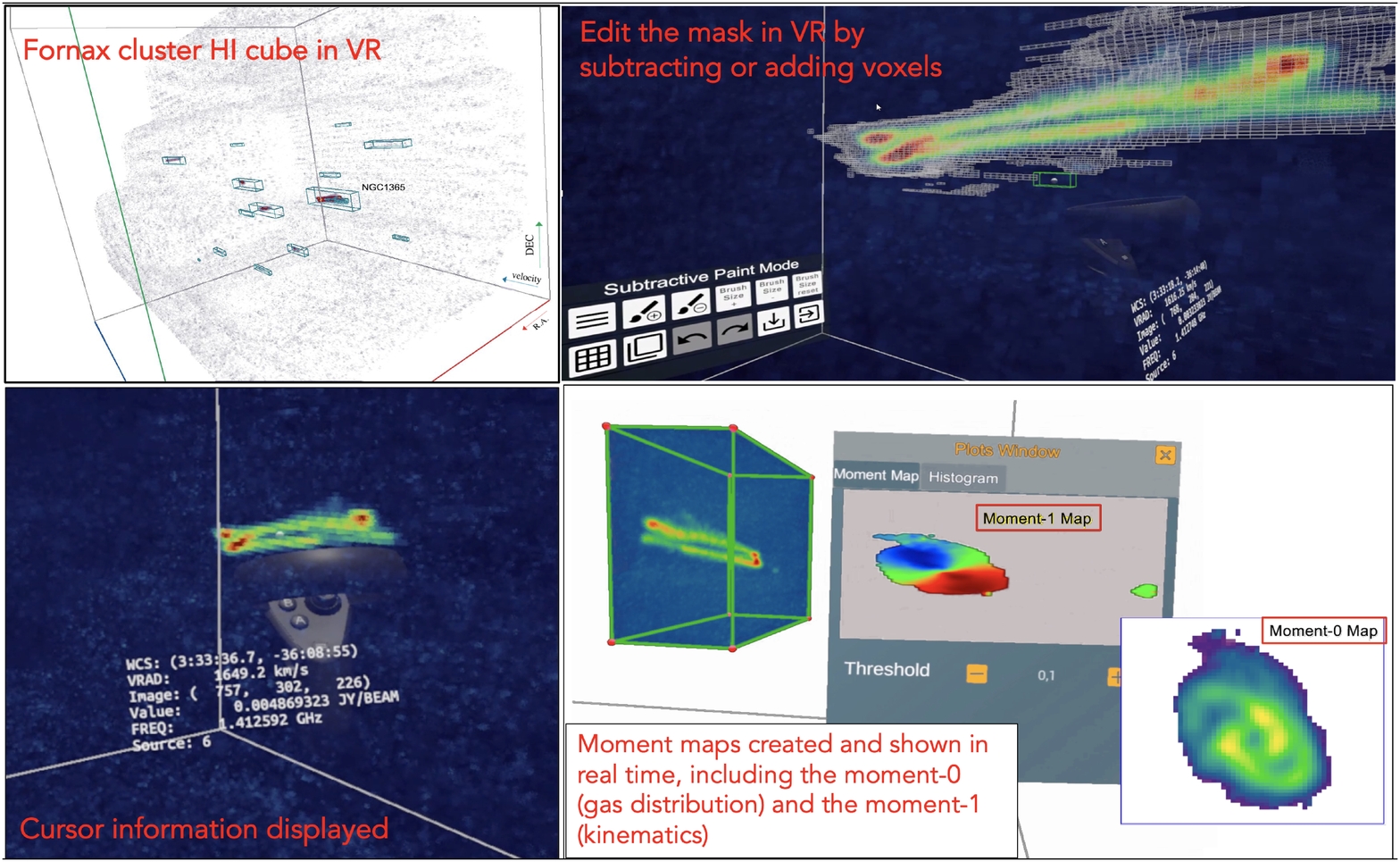}{fig:plotVR}{Some \texttt{iDaVIE-v} capabilities. \textbf{Top left:} \texttt{iDaVIE-v} VR environment view of the H{\textsc i} Australian Compact Telescope Array (ATCA) cube of the Fornax cluster (courtesy  of P. Serra and collaborators, \citealt{Serra16}). The sources are highlighted with cyan boxes. \textbf{Bottom left:} when the hand controller cursor intercepts an identified source in the cube, key source information extracted from the cube and mask (if any) are displayed to the user. \textbf{Top right:} illustration of \texttt{iDaVIE-v} in "edit mask" mode; in this case the user is deleting misidentified voxels from the mask shown as a grey voxels grid on the data. \textbf{Bottom right:} illustration of moment maps rendering of a selected source, NGC\,1365, a large barred spiral in the Fornax Cluster}
%; \citep{Maddox19}} 
\item \textbf{Data export.} Once an operation is completed the results can be exported and saved to disk. 
The outputs depend on the operations the user carried out on the data.
There are several outputs that are originated by \texttt{iDaVie-v} such as plots, moment maps, screenshots, an amended mask and a new source catalogue. 
\end{itemize}

\section{\texttt{iDaVIE-v} beta release \& future development}
The beta release of \texttt{iDaVIE-v} is expected for early 2021. The release will include the \texttt{iDaVIE-v} executable file and user instructions. The source code will be made publicly available only when the final release will be completed (late 2021). \texttt{iDaVIE-v} development will continue after the beta release in order to include other functionalities (such as the capability of importing and visualise multiple catalogues in one go) and will also rely on feed-backs from the user community for further improvement suggestions. %Final goal of the IVL work is to integrate all the \texttt{iDaVIE} software suite modes (\texttt{iDaVIE-v}, \texttt{iDaVIE-p} anf \texttt{iDaVIE-d}; \citet{P4-6_adassxxviii}) in a single (possibly internet based) software tool that can 

\bibliography{O3-130}

%%%%%%%%%%%%%%%%%%%%%%%%%% END %%%%%%%%%%%%%%%%

% if we have space left, we might add a conference photograph here. Leave commented for now.
% \bookpartphoto[width=1.0\textwidth]{foobar.eps}{FooBar Photo (Photo: Any Photographer)}

\end{document}